\documentclass[iop]{emulateapj}
\usepackage{graphicx}
\usepackage{bm}% bold math
\usepackage{color}
\usepackage{subfigure}
\usepackage{epstopdf}
\usepackage{wasysym}
\usepackage{hyperref}
\usepackage{url}

\def\mnras{MNRAS}
\def\apj{ApJ}

\def\simleq{\; \raise0.3ex\hbox{$<$\kern-0.75em \raise-1.1ex\hbox{$\sim$}}\; }
\def\simgeq{\; \raise0.3ex\hbox{$>$\kern-0.75em \raise-1.1ex\hbox{$\sim$}}\; }

\newcommand{\GeV}{{\rm GeV}}
\newcommand{\TeV}{{\rm TeV}}

\newcommand{\kpc}{{\rm kpc}}

\newcommand{\km}{{\rm km}}

\newcommand{\s}{{\rm s}}
\begin{document}

\title{The gamma-ray and neutrino sky: \\ a consistent picture of Fermi-LAT, Milagro, and IceCube results}

\author{Daniele Gaggero}
\affil{SISSA and INFN, via Bonomea 265, I-34136 Trieste, Italy}
\affil{GRAPPA Institute, University of Amsterdam, Science Park 904, 1090 GL Amsterdam, The Netherlands}
\email{d.gaggero@uva.nl}

\author{Dario Grasso}
\affil{INFN and Dipartimento di Fisica ``E. Fermi", Pisa University, Largo B. Pontecorvo 3, I-56127 Pisa, Italy}
\email{dario.grasso@pi.infn.it}

\author{Antonio Marinelli}
\affil{INFN and Dipartimento di Fisica ``E. Fermi", Pisa University, Largo B. Pontecorvo 3, I-56127 Pisa, Italy}
\email{antonio.marinelli@pi.infn.it}

\author{Alfredo Urbano}
\affil{SISSA and INFN, via Bonomea 265, I-34136 Trieste, Italy}
\affil{CERN, Theory division, CH-1211 Gen\`eve 23, Switzerland}
\email{alfredo.leonardo.urbano@cern.ch}

\author{Mauro Valli}
\affil{SISSA and INFN, via Bonomea 265, I-34136 Trieste, Italy}
\email{mauro.valli@sissa.it}

\begin{abstract}

We compute the $\gamma$-ray and neutrino diffuse emission of the Galaxy on the basis of a recently proposed phenomenological model characterized by radially dependent cosmic-ray (CR) transport properties. We show how this model, designed to reproduce both {\it Fermi}-LAT $\gamma$-ray data and local CR observables, naturally reproduces the anomalous TeV diffuse emission observed by Milagro in the inner Galactic plane. 
Above 100 TeV our picture predicts a neutrino flux that is about five (two) times larger than the neutrino flux computed with conventional models in the Galactic Center region (full-sky). Explaining in that way up to $\sim 25 \%$ of the flux measured by IceCube, 
we reproduce the full-sky IceCube spectrum adding an extra-Galactic component derived from the muonic neutrino flux in the northern hemisphere. We also present precise predictions for the Galactic plane region where the flux is dominated by the Galactic emission.
%Our results may help to explain hints of an excess of HESE events in the southern hemisphere along the Galactic Plane, as well as those of different spectral slopes in the northern and southern hemispheres found by IceCube.  
%Our predictions will be testable in the near future by neutrino observatories in the northern hemisphere such as ANTARES and  KM3NeT as well as by IceCube dedicated analyses focused on the Galactic plane.

\end{abstract}

\maketitle 

\section{introduction.}

In recent years the IceCube collaboration has opened the era of neutrino astronomy and announced the detection of $37$ extraterrestrial neutrinos above $\sim 30~\TeV$~\citep{Aartsen:2013bka, Aartsen:2013jdh, Aartsen:2014gkd}. 
% corresponding to an excess of $5.7~\sigma$ with respect to the atmospheric background
%The inferred flavor composition is compatible with a mixture of electronic, muonic and tauonic neutrino in equal amounts as expected if their origin were astrophysical~\citep{Aartsen:2015nda}. 
More recently, a preliminary analysis~\citep{IC_4yr_ICRC}, based on four years of data, rose the total number of high-energy starting events (HESE) to $54$.

{  
The astrophysical spectrum inferred by IceCube on the basis on the three-year data set was fitted by a power law with index $\Gamma = - 2.3 \pm 0.3$ above 60 TeV~\citep{Aartsen:2014gkd}, while the four-year data favor a steeper spectrum: $\Gamma = - 2.58 \pm 0.25$ (\citep{IC_4yr_ICRC}).
Although a statistically significant departure from isotropy cannot be claimed yet, a recent analysis~\citep{Ahlers:2015moa} showed that the angular distribution of HESE events allows up to $50\%$ of the full-sky astrophysical flux to have a Galactic origin. Moreover, a hint of a harder spectrum in the northern Hemisphere may be suggested by a recent analysis~\citep{Aartsen:2015rwa}.
}

The Galaxy, indeed, is a guaranteed source of neutrinos up to a fraction of PeV energies at least.

A sizable flux may either come from freshly accelerated cosmic rays (CRs) undergoing hadronic scattering with gas clumps, or from the hadronic interactions between the Galactic CR sea and the diffuse gas.

The former scenario, however, cannot explain the steepness of the neutrino spectrum measured by IceCube and is in tension with {\it Fermi}-LAT upper limit on the corresponding $\gamma$-ray emission~\citep{Tchernin:2013wfa}.

{  In the latter, instead, if the local CR spectrum is assumed to be representative of the entire Galactic population, the computed spectrum should be significantly lower than the measured spectrum~\citep{Stecker:1978ah,Berezinsky:1992wr,Evoli:2007iy} %(unless an unrealistically high gas density is assumed: see~\citep{WW}) 
(see also~\citep{Ahlers:2015moa}: The authors show that only $\simeq 8\%$ of the HESE can be accounted in that way, under the conventional assumption that the same CR transport properties hold throughout the whole Galaxy).}
 
%That model was tuned on updated CR data as well as on the diffuse $\gamma$-ray emission of the Galaxy measured by Fermi-LAT~\citep{FermiLAT:2012aa} and assumes that the same CR transport properties hold in whole Galaxy ({\it conventional scenario}).  

However, it is conceivable that CR diffusion -- due to a stronger star forming activity and peculiar magnetic field strength/geometry -- behaves differently in the inner Galactic region. 
Several anomalies observed in the $\gamma$-ray diffuse emission support this possibility.

\begin{figure}[tbp]
\centering
\includegraphics[width=0.45\textwidth]{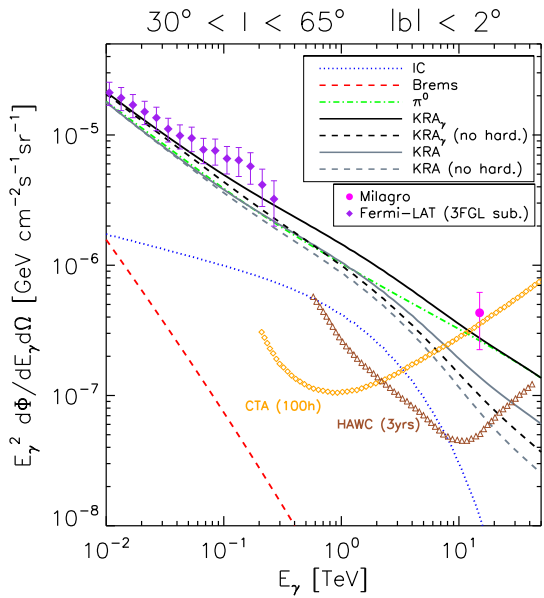}
\caption{Diffuse emission $\gamma$-ray spectrum from the inner
Galactic plane ($|b| < 2^\circ$,   $30^\circ < l < 65^\circ$) computed for the
reference models considered in this Letter is compared with
{\it Fermi}-LAT and Milagro data. The Milagro differential flux reported
here is 17\,\% lower with respect to the flux reported in 2008 (\citep{MilagroPlane}) due to the assumption
of a spectral index of 2.4 instead of 2.7 (\citep{MilagroPrivate}).
The expected sensitivities of HAWC~\citep{HAWK} and CTA \citep{CTA} are reported. The spectral components are shown for the KRA$_\gamma$ model only. {  The {\it Fermi}-LAT data points refer to 5 years of data, within the event class ULTRACLEAN, according to {\it Fermi} tools v9r32p5.}}
\label{fig:milagro}
\end{figure}

We start by noting that conventional models cannot explain the large $\gamma$-ray flux measured by the Milagro observatory from the inner GP region at $15~\TeV$ median energy~\citep{Beacom2007,Abdo:2008if}.  
In Fig.~\ref{fig:milagro} we show how a representative conventional model, with similar spectral properties as the {\it Fermi} benchmark model \citep{FermiLAT:2012aa}, clearly fails to reproduce that measurement.
{This problem is common to all the models of this kind and still holds assuming -- as done in~\citep{Ahlers:2015moa} --  that the spectral hardening} found by PAMELA in the CR proton and helium spectra above $\sim 230~\GeV/{\rm n}$~\citep{Adriani:2012paa} is present throughout the whole Galaxy.
Therefore, the Milagro excess is still an open issue, and indeed its possible relevance for high-energy neutrino physics has often been pointed out (see e.g.~\citep{Gabici:2008gw,Taylor:2014hya}). 

An even more serious anomaly was found at lower energies in the {\it Fermi}-LAT diffuse $\gamma$-ray spectrum~\citep{FermiLAT:2012aa}: {the conventional models systematically underestimate the measured flux in the inner GP region above a few GeV.}
A new phenomenological scenario was proposed in~\citep{Gaggero:2014xla} in order to account for the latter results: the idea is to consider a radial dependence for both the rigidity scaling index $\delta$ of the diffusion coefficient and the {  advective} wind.

In this Letter we present for the first time a consistent picture based on that scenario that aims to overcome all of the aforementioned problems. 

%This is based on the KRA$_\gamma$ model assuming spatial dependent CR transport to explain Fermi-LAT results. 
%CR propagation aiming to overcome all the aforementioned problems, exploring the possibility of radial gradients in CR transport. The model we propose here, originally motivated in~\citep{Gaggero:2014xla}, is properly tuned in order to take into account very-high-energy proton and Helium local measurements.
The most significant achievements we present are:
\begin{itemize}
\item a natural explanation to the long-standing Milagro anomaly;
\item a new prediction of the Galactic neutrino diffuse emission that is significantly larger than the one computed with conventional models; and
%spectrum measured by IC, with a larger contribution of Galactic neutrinos to the total flux;
%\item  for the neutrino flux in the Galactic plane region; , testable in the near future by neutrino observatories with a dedicated analysis;
\item a possible interpretation for the hints of an excess of IceCube events along the Galactic plane and of the different neutrino slope in the northern and southern hemispheres.% This is also testable in the near future, with more accurate data.
\end{itemize}

% We remark here such a scenario is testable in foreseeable dedicated analyses of neutrino observatories concerning the Galactic plane region.
  
\section{The model.}

{ 
Following \citep{Gaggero:2014xla}, the starting point  is a conventional propagation setup characterized by $\delta = 0.5~$\footnote{$\delta$ is defined by $D(\rho) \propto \left(\rho/\rho_0\right)^\delta$}, compatible with a Kraichnan spectrum of the interstellar turbulence within the quasi-linear theory framework. 
We will refer to this setup as the ``KRA model'' (see also~\citep{Evoli2011id}). 

The new model presented in that paper features $\delta$ {increasing} with the galactocentric radius $R$ (implying spatially variable CR transport as originally suggested, e.g., in~\citep{Erlykin:2012dp}), and hence predicts a hardening of CR propagated spectrum and $\gamma$-ray emissivity in the inner Galaxy. 
The following explains the model in more detail:

\begin{itemize}
\item $\delta$ has the galactocentric radial dependence $\delta(R) = A R + B$ for $R < 11~\kpc$ where $A = 0.035~\kpc^{-1}$ and $B = 0.21$ so that $\delta(R_{\odot}) = 0.5$. This behavior may have different physical interpretations, e.g. a smooth transition between a dominant parallel escape along the poloidal component of the regular Galactic magnetic field (in the inner Galaxy, where $\delta$ is lower) and a perpendicular escape with respect to the regular field lying in the plane (in the outer Galaxy, where the scaling is steeper). 

\item An { advective} wind for $R < 6.5~\kpc$ with velocity $V_C(z) {\hat z}$ ($z$ is the distance from the GP) vanishing at $z = 0$ and growing as $dV_c/dz = 100~\km~ \s^{-1}~ \kpc^{-1}$ is also included. This ingredient is motivated by the X-ray
ROSAT observations~\cite{rosat}

\item The vertical dependence of the diffusion coefficient is taken as $D(z) \propto \exp(z/z_t)$;  

\item The halo size is $z_t = 4~\kpc$ for all values of $R$ (this is a conventional choice widely used in the literature, and we checked that our results do not change significantly considering larger values of $z_t$).
\end{itemize}

The observed $\gamma$-ray spectra at both low and mid Galactic latitudes, including the Galactic center, are reproduced by this model without spoiling local CR observables: proton, antiproton and helium spectra, B/C and $^{10}$Be/$^{9}$Be ratios. Moreover, this scenario naturally accounts for the radial dependence in the CR spectrum found by the Fermi collaboration~\citep{Casandjian:2015ura}.  
We will refer to this model as ``KRA$_{\gamma}$'' since it is tuned on gamma-ray data.
}

We implement the setup with {\tt DRAGON}, a numerical code designed to compute the propagation of all CR species~\citep{Evoli:2008dv, Dragonweb}.  {  While the current version of the code shares with {\tt GALPROP}~\citep{Galpropweb} the same spallation cross-section routines and the gas distribution, its innovative structure allows us to compute CR transport in the general framework of position-dependent diffusion. }

%The model parameters were tuned to reproduce the slope and angular distribution of the diffuse $\gamma$-ray emission of the Galaxy measured by Fermi-LAT without spoiling the main local CR observables: {proton, antiproton and Helium spectra, B/C and $^{10}$Be/$^{9}$Be ratios.}
%
Concerning the $p$ and He spectral hardening inferred from PAMELA~\citep{Adriani:2011cu} - recently confirmed by AMS-02~\citep{Aguilar:2015ooa} -
and CREAM~\citep{Ahn:2010gv} data above $\sim 250~\GeV/{\rm n}$, we consider two alternatives. {\it 1) Local hardening} could originate from nearby supernova remnants (see e.g.~\citep{Thoudam:2013sia}); since this is a stochastic effect and averages out on large scales it amounts to not introducing any feature in the Galactic CR population used in this work. {\it 2) Global hardening} could originate from a spectral feature in the rigidity dependence of CR source spectra or the diffusion coefficient (here we only consider the former case, as both scenarios have the same effect on the $\gamma$-ray diffuse emission).
In both cases we assume that above 250 GeV/n the CR source spectra extend steadily up to an exponential cutoff at the energy $E_{\rm cut}/{\rm nucleon}$.  

We consider two representative values of this quantity, namely $E_{\rm cut} = 5$ and $50$ PeV which - for the KRA$_\gamma$ setup -
match CREAM {\it p} and He data and roughly bracket KASCADE~\citep{KASCADE2005} and KASCADE-Grande data~\citep{Apel:2013dga}. 
While KASCADE proton data favor the lowest cutoff, the highest is favored by the KASCADE-grande all-particle spectrum.
A more detailed fit of the CR spectra in the PeV region is not justified here due to the large experimental uncertainties on the elemental spectral shapes and normalizations. The consequent uncertainty on the neutrino flux should, however, be well captured by our choice to consider a range of cutoffs.

\section{The $\gamma$-ray spectrum.}  

As shown in~\citep{Gaggero:2014xla}, the KRA$_\gamma$ setup --  both in its  {\it local} (KRA$_{\gamma}$ with no hardening) and {\it global} realizations -- provides a good fit of the $\gamma$-ray diffuse emission measured by {\it Fermi}-LAT all over the sky, particularly toward the inner GP region. 
Moreover, it accounts for the galactocentric radial dependence of the CR spectral index found by the {\it Fermi}-LAT collaboration~\citep{Casandjian:2015ura}.

Here we extend the computation performed in \citep{Gaggero:2014xla} above the TeV. % to test its against Milagro results. 

Similar to~\citep{Gaggero:2014xla}, {we compute the hadronic emission integrating the expression of the $\gamma$-ray emissivity along the line of sight using {\tt GammaSky}, a dedicated code used in~\citep{gradient:2012, Cirelli2014} to simulate diffuse $\gamma$-ray maps. This package features, among other options, the gas maps included in the public version of {\tt GALPROP}~\citep{Galpropweb,Moskalenko2002,FermiLAT:2012aa}.
We adopt the emissivities given in~\citep{kamae}, accounting for the energy dependence of the $pp$ inelastic cross section (significant above the TeV).}
We disregard $\gamma$-ray opacity due to the interstellar radiation field, since it is negligible up to a few tens of TeV~\citep{Ahlers:2013xia}. 

Our results are shown in Fig.~\ref{fig:milagro}.
As mentioned in the above, a representative conventional model (KRA) cannot account for the flux measured by Milagro from the inner GP at $15$ TeV even if accounting for the CR spectral hardening required to match the PAMELA and CREAM data.
The KRA$_\gamma$ setup, instead, is more successful, especially if a global hardening is assumed.
This is a remarkable result since: {\it 1)} it {  supports the KRA$_\gamma$ model} in a higher-energy regime; {\it 2)} it provides the first consistent interpretation of Milagro and {\it Fermi}-LAT results (an {\it optimized} model was proposed to account for the EGRET GeV excess \cite{Strong:2004de}, and came out to reproduce Milagro results as well, but was subsequently excluded by {\it Fermi}-LAT  \cite{Abdo:2009mr}), and {\it 3)} it reinforces the arguments in favor of a non-local origin of the hardening in the CR spectra above 250 GeV. 

{  Interestingly, the KRA$_\gamma$ model also reproduces the high-energy diffuse $\gamma$-ray spectrum measured by H.E.S.S. in the Galactic ridge region ($| l | < 0.8^\circ$,  $| b | < 0.3^\circ$) in terms of CR scattering with the dense gas in the central molecular zone without the need to invoke the contribution of sources to that region~\citep{Aharonian:2006au} and without further tuning (see ~\citep{Gaggero:2015jma} for more details).} Although this is a very small region with respect to the regions considered in this paper,  this result may be interpreted as a valuable check of our model in a region not covered by Milagro 
%{  (see also \citep{Candia} for an alternative scenario leading to an enhanced gamma-ray and neutrino flux from the GC, as a consequence of peculiar transport properties). }

{  Moreover, our KRA$_{\gamma}$ model is also compatible with ARGO-YBJ results in the window $65^\circ < l < 85^\circ$ and $|b| < 5^\circ$; both the KRA and the KRA$_{\gamma}$ are consistent with CASA-MIA measurements at  high Galactic longitudes~\citep{casamia}.}
        
\section{The neutrino emission.}

The results discussed above clearly show that the hadronic emission computed with the KRA$_\gamma$ setup above the TeV is significantly stronger than the conventional model predictions. In this section we show the relevant consequences concerning the Galactic neutrino emission.

We first compute the $\nu_e$ and $\nu_\mu$ production spectra: for both flavors we use the emissivities provided in~\citep{kamae} (well tuned on accelerator and CR data) for projectile energies below $\sim 500$ TeV, while we adopt the emissivities provided in \citep{Kelner:2006tc} that are above that energy range.
Then we account for neutrino oscillations: their effect is to almost equally redistribute the composition among the three flavors~\citep{Cavasinni:2006nx}.  
We only consider proton and helium CRs/gas -- as for $\gamma$-rays -- since heavier nuclear species give a negligible contribution in the energy range we cover in this work~\citep{Kachelriess:2014oma}. 

\begin{figure}[htb!]
\centering
\includegraphics[width=0.45\textwidth]{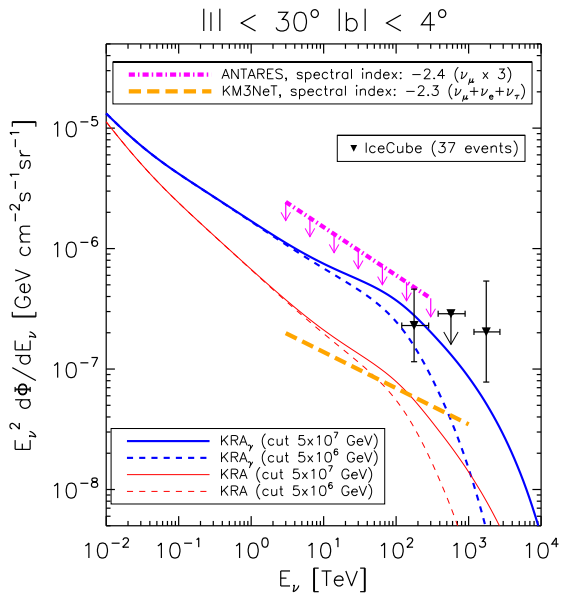}  
\caption{ Solid and dashed red (blue) lines: expected neutrino spectra {  (all flavors, both neutrinos and antineutrinos)} in the inner Galactic plane region computed for the conventional KRA (the novel KRA$_\gamma$) models for two different cutoff values.
%Dotted and dash-dotted lines: Combination of Galactic and extragalactic spectra for the same models. 
%The EG flux is inferred from the northern hemisphere astrophysical muon neutrino IC measurements~\citep{Aartsen:2015rwa}.
We also show the maximal flux, estimated considering three years of IceCube HESE events as described in \citep{Spurio:2014una}, the constraint from the ANTARES experiment ~\citep{TheANTARES:2015pba}  (1500 days of experiment livetime between 2007 and 2013) as well as the deduced sensitivity of the future Mediterranean observatory KM3NeT~\citep{KM3NeT-Piattelli:2015} with four years ($\sim$ 1500 days) of livetime. 
}
\label{fig:nu_gal_plane}
\end{figure}

Because neutrinos in the Galactic emission are expected to be maximal in the inner Galactic plane region, we first present our results for the window $| l |<30^{\circ}$ and $| b |<4^{\circ}$.
For this region the ANTARES collaboration~\citep{Antares} recently released an upper limit on the muon neutrino flux based on the result of an unblinding analysis regarding the events collected between 2007 and 2013 in the energy range [$3 \div 300$] TeV~\citep{TheANTARES:2015pba}. 

In Fig.~\ref{fig:nu_gal_plane} we compare the $\nu_\mu$ flux computed with the KRA and KRA$_\gamma$ setups with the flux of the experimental constraint.  
First of all we notice the large enhancement (almost a factor of 5 at 100 TeV) obtained with the KRA$_\gamma$ model with respect to the conventional scenario. 
Indeed, while -- in agreement with previous results -- we find that the flux corresponding to the KRA model may require long times of observation even by the KM3NeT observatory~\citep{Nemo}, our prediction for the KRA$_\gamma$ model is instead well above the sensitivity reachable by that experiment in four years and it is almost within the ANTARES observation capabilities.
 
\begin{figure}[tbp]
    \begin{center}
%\vspace{-0.5cm}
\includegraphics[width=0.45\textwidth]{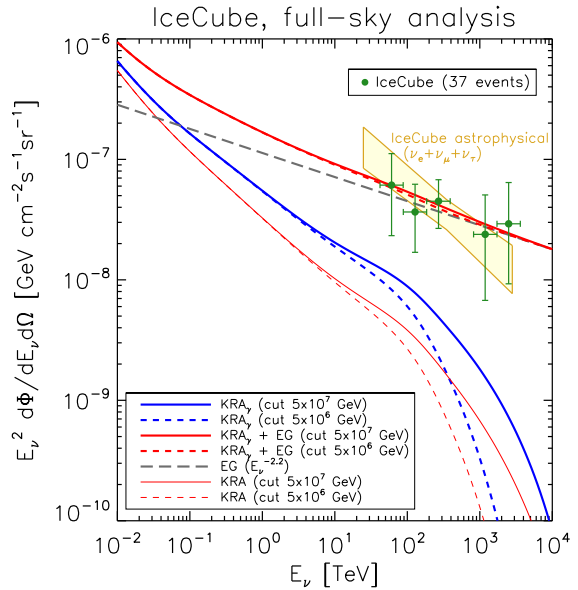}
\caption{Full-sky neutrino spectrum {  (all flavors, both neutrinos and antineutrinos)} predicted by the KRA$_\gamma$ and KRA models (with global CR hardening), adopting two different choices for the CR high-energy cutoff. We also plot the combination of the Galactic (KRA$_\gamma$) and a benchmark EG spectrum. The EG flux is consistent with that inferred from the IceCube collaboration in the northern hemisphere~\citep{Aartsen:2015rwa}.
The models are compared with the 68\% confidence region for the IceCube astrophysical neutrino flux obtained with a maximum-likelihood (yellow region) ~\citep{Aartsen:2015knd} and the three years HESE (green points)~\citep{Aartsen:2014gkd}.
}
%The enhanced KRA$_\gamma$ Galactic contribution clearly improves the fit to the data.}
\label{fig:nu_spectra}
\end{center}
\end{figure}

Interestingly, our result is in good agreement with the maximal flux inferred from the fraction of IceCube HESE events compatible with that region (see Fig.~\ref{fig:nu_spectra}).
We notice that in that region the expected EG contribution, as constrained from the muon neutrino flux in the northern hemisphere (see below) gives a subdominant contribution with respect to that computed with the KRA$_\gamma$ model. Therefore the possible detection of a signal in that sky window would be a smoking gun for the presence of such Galactic emission.  

IceCube should also have the potential to detect that emission on a larger region. 
%We checked that the neutrino flux computed with the KRA$_\gamma$ model for $|b| < 7.5^o$ is in good agreement with that inferred from IceCube HESE analysis.
In this context, we also note that an independent analysis \citep{Neronov:2015osa} already found a significant hint of an excess in the 4-year HESE sample~\citep{IC_4yr_ICRC} along the Galactic plane.

We now turn our attention to the recently published IceCube results, both concerning the full-sky and the northern and southern hemispheres separately.   

In Fig.~\ref{fig:nu_spectra} we represent the full-sky total neutrino spectrum (all flavors, including antiparticles) computed for the KRA$_\gamma$ and KRA models, {with} global CR hardening, and compare it to the IceCube results.  

Our prediction for the conventional setup (KRA model) is in good agreement with~\citep{Ahlers:2015moa}: In that work, the benchmark Galactic model accounts for $8$\% of the flux measured by IceCube above $60$ TeV, for a CR spectrum similar to the one used here above $50$ PeV.

On the other hand, the KRA$_\gamma$ predicts a $\sim 2$ times larger full-sky flux above $10$ TeV: the model prediction is therefore only $\simeq 4$ times smaller than the best fit of the astrophysical flux measured by IceCube on the whole sky. 

{  We remark that another  analysis~\citep{Neronov2015app}, based on an extrapolation of {\it Fermi}-LAT data, points toward a non-negligible Galactic contribution to the full-sky neutrino flux due to a hard diffuse CR spectrum. In that scenario the (softer) locally observed CR spectrum may get a major contribution from one or more local sources: this interpretation still has to be validated against {\it Fermi}-LAT data, while our model is based on those measurements.} 

Setting a threshold energy at $60$ TeV and convolving the KRA$_\gamma$ spectrum (with $E_{\rm cut} \,=\, 50$ PeV) with the IceCube HESE effective areas~\citep{Aartsen:2013jdh}, the expected number of neutrino events in three years of IceCube observation represents $\sim15\%$ of the published sample~\citep{Aartsen:2014gkd}. These rates are well above those expected due to atmospheric muons and atmospheric neutrinos and confirm the spectral comparison between KRA$_\gamma$ and IceCube data.

Clearly, another component -- most likely of extragalactic (EG) origin -- needs to be invoked in order to account for all of the IceCube events.

{ 
Here we assume this EG component to be isotropic and use the astrophysical muon neutrino IceCube measurements from the northern hemisphere~\citep{Aartsen:2015rwa} -- where the Galactic emission is only $\sim 1/10$ of the total flux --  to probe its spectral properties.
%Fitting that spectrum with a single power law the IceCube collaboration found  $\Phi^{\rm North}_{\nu_\mu} = 1.7^{+ 0.6}_{- 0.8} \times 10^{-18}~\left(E/100~{\rm TeV}\right)^{-2.2 \pm 0.2}~(\GeV~ \cm^{2}~ \sr~ \s)^{-1}$ which, under the assumption of universal flavor composition, should be multiplied by three to get the total neutrino flux in the northern hemisphere.  
Although the northern spectral slope is statistically compatible with the full-sky one, % we remind that, as mentioned in the Introduction, both the preliminary spectrum inferred from the 4-year set of $54$ HESE \citep{IC_4yr_ICRC} and the results of~\citep{Aartsen:2015knd} provide an interesting hint of a steeper ($\Gamma \simeq - 2.5$) and larger spectrum in the southern hemisphere.
given the hint of a steeper spectrum in the southern hemisphere, it is interesting to check if the combination of our Galactic prediction and the EG flux inferred from the aforementioned muon neutrino measurement provide a better agreement with the data. %if the combination of our Galactic prediction and the EG flux inferred from the aforementioned muon neutrino measurement provides a better agreement of the full-sky flux, although a statistical analysis is still premature due to the large uncertainties.
}

For illustrative purposes, in Fig.~\ref{fig:nu_spectra} we show the effect of adding an isotropic EG emission to the Galactic neutrino emission computed with the KRA$_{\gamma}$ model, with a spectrum given by the IceCube best fit of $\Phi^{\rm North}_{\nu_\mu}$, multiplied by three to account for all flavors. {  The nature of such emission is still under debate: as pointed out in \citep{Glusenkamp:2015} and \citep{Becktol:2015}, neither blazars nor star-forming galaxies can provide more than a subdominant contribution, given the constraints imposed by the gamma-ray extragalactic background inferred from {\it Fermi}-LAT data}. 
{  The plot clearly shows how the KRA$_{\gamma}$ helps to improve the fit in the low-energy part of the IceCube spectrum.

We also checked that the neutrino flux computed with the KRA$_\gamma$ model for $|b| < 7.5^o$ is in rather good agreement with that inferred from IceCube HESE analysis if the EG emission, as estimated above, is accounted for.  A dedicated analysis will be performed in a forthcoming work.}
%Such an emission is compatible with that expected by known extra-Galactic sources, e.g. by starburst galaxies~\citep{Loeb:2006tw}.   

\section{Conclusions.}
 
 In this Letter we connected $\gamma$-ray GeV and TeV measurements in a unified scenario, together with the recently released IceCube neutrino data, providing a consistent picture based on a CR transport model proposed in~\citep{Gaggero:2014xla}. 
The model features a variation of the diffusion coefficient rigidity scaling $\delta$ with galactocentric radius. The variation was suggested by a spectral anomaly found in the {\it Fermi}-LAT $\gamma$-ray data, and turned out to be compatible with both $\gamma$-ray spectra at low and intermediate Galactic latitude and local CR observables. 

In this work we showed that our picture sheds new light on high-energy gamma-ray and neutrino recent observations.
In particular, it provides a novel natural explanation for the anomalous $\gamma$-ray flux measured by the Milagro observatory from the inner GP region at $15~\TeV$; moreover, it appears to be compatible with the H.E.S.S. spectrum in the Galactic ridge region.

Remarkably, our model also provides a different interpretation of the full-sky neutrino spectrum measured by IceCube with respect to the standard lore, since it predicts a larger contribution of the Galactic neutrinos to the total flux, compared to conventional models. 

These predictions will be testable in the near future by neutrino observatories such as ANTARES, KM3NeT, and IceCube itself via dedicated analyses that are focused on the Galactic plane, and also by analyzing the different spectral slopes in the northern and southern hemispheres. 
A hint of a softer slope in the northern hemisphere is already present, and appears to be compatible with our picture.

A physical interpretation of our model most likely requires either abandoning the isotropic diffusion scenario generally adopted to treat CR propagation, or considering different turbulence regimes in different regions of the Galaxy: a quantitative modeling of those phenomena is far beyond the scope of our phenomenological work. %Forthcoming measurements by IceCube and KM3NeT as well as $\gamma$-ray observations by HAWC and CTA will have the potentiality to test the picture we propose. 

\vskip 0.2 cm

\noindent{\underline{\em Acknowledgments}:} We are indebted to Carmelo Evoli, Piero Ullio, Andrii Neronov, Petra Huentemeyer, Markus Ahlers, Maurizio Spurio, Luigi Fusco, and Rosa Coniglione for many inspiring discussions and useful advice.

D.\,Gaggero acknowledges the SFB 676 research fellowship from the University of Hamburg as well as the hospitality of DESY. 
A.\,Marinelli acknowledges the Galilei research fellowship of Pisa University.
The work of A.U. is supported by the ERC Advanced Grant n$^{\circ}$ $267985$, ``Electroweak Symmetry Breaking, Flavour and Dark Matter: One Solution for Three Mysteries" (DaMeSyFla).

\vskip 0.2 cm

\bibliographystyle{myapj}
%\bibliographystyle{apsrev4-1}
%\bibliographystyle{apj_hyperref}
%\bibliography{bibneutrino}

\end{document}